# CanariCam Mid-Infrared Drift Scanning: Improved Sensitivity and Spatial Resolution

Amílcar R. Torres-Quijano[1], Christopher Packham[1,2,3], and Sergio Fernandez Acosta[4]

[1] Department of Physics & Astronomy, University of Texas at San Antonio, 1 UTSA Circle, San Antonio, Texas, 78249, USA
[2] National Astronomical Observatory of Japan, 2-21-1 Osawa, Mitaka, Tokyo 181-8588, Japan
[3] NASA Marshall Space Flight Center, Huntsville, Alabama, United States
[4] GranTeCan, Instituto de Astrofísica de Canarias, C/ Vía Láctea, s/n E-38205 La Laguna (Tenerife), Spain

E-mail: amilcar.torres-quijano@utsa.edu

## Abstract

Ground-based mid-infrared (MIR) astronomical observations require the removal of the fast time variable components of (a) sky/background variation and (b) array background. Typically, this has been achieved through oscillating the telescope's secondary mirror a few times a second, a process termed "chopping". However, chopping reduces on-object photon collection time, imposes stringent demands on the secondary mirror, requires nodding of the telescope to remove the radiative offset imprinted by the chopping, and relies on an often-fixed chop-frequency regardless of the sky conditions in the actual observations. In the 30m telescope era, secondary mirror chopping is impracticable. However, if the sky and background is sufficiently stable, drift scanning holds the promise to remove the necessity of chopping. In this paper we report our encouraging drift scanning results using the CanariCam MIR instrument on the 10.4m Gran Telescopio Canarias and the implications to future instruments and experiments.

## 1. Introduction

The technique of 'chopping' in thermal-infared (TIR, ~5-26 μm) astronomy is well established. The sky is not dark at infared (IR, ~1-26 μm) wavelengths, and both the telescope and instrument emit copious numbers of TIR photons. Indeed, the TIR background emission can be many orders of magnitudes greater than the emitted source photons, necessitating specialized techiques to measure and correct the background. Chopping serves to first image the source and background, and then oscillate the mirror to a 'blank' sky area, where the background can be measured. This works well if the oscillation of the mirror is faster than the background changes. Thus a chopping mirror, typically the telescope secondary, is oscillated at several Hertz (Hz). It is important to remember that whilst this can effectively correct for the typically very high level of background photons, it does not remove the statistical noise of these photons, rather it actually increases the variance by a factor of two. Further, chopping (and the required nodding, see §1.2) induces several deleterious effects, opto-mechanical requirements, and efficiency issues, and is thus far from a zero-cost solution. In this paper, we build and update our work published by Packham, Torres-Quijano, and Fernandez Acostsa 2021 (PTF 2021 hereafter), detailing how we have refined our data reduction approach, leading to a much improved final result. We first discuss the history of the CanariCam MIR instrument on the Gran Telescopio Canarias (GTC), chopping and nodding, and an overview of drift scanning (§1). In §2 we discuss our experimental design and observing log, and in §3 we detail our data reduction approach and steps. In §4 we detail our results and in §5 we make our conclusions and a brief discussion.

## 1.1 History of CanariCam

CanariCam (Telesco et al., 2003) was the facility multi-mode MIR (mid-infrared (IR), ~7.5-25 μm) camera on the 10.4 m GTC on La Palma, Spain. It was designed and built by the University of Florida (PI: Charles Telesco) and afforded the GTC community imaging, spectroscopic and unique polarimetric MIR capabilities with a full width at half maximum (FWHM) at, or near, the diffraction limit of the telescope, but the Strehl ratio was typically ~30%[1] in the *N* band atmospheric window (~7.5-13 μm). Since 2012, it operated in queue mode at a Nasmyth focal station, until it was temporarily decommissioned in 2016. Following an upgrade project (Fernandez-Acosta, 2020) started in 2018, it was reinstalled and recommissioned in a folded-Cassegrain focal station in late 2019, retrofitted with a more powerful and reliable 4 Kelvin cryocooler and new electronics and interfaces. In January 2021 the instrument was finally decommissioned after 9 years of service to the GTC community.

Due to the versatile instrumental modes, large collecting area of the GTC, and high spatial resolution, CanariCam served a broad range of science cases in fields spanning Active Galactic Nuclei (i.e. Alonso-Herrero et al. 2016), supernovae (i.e. Telesco et al. 2015), the Galactic Center (Roche et al. 2018), protoplanetary disks (i.e. Li et al. 2016) and substellar objects among others, and its polarimetric modes (Packham et al. 2005) were unique, enabling the study of magnetic fields through the absorption or emission of aligned dust grains. It has produced 46 refereed papers, with 607 citations and an H-Index of 15, from a total of 876 hours of telescope time. After removing the hours delivered during the last year (in accordance with the GTC metrics to account for publication lag), it corresponds to a good productivity ratio of 18 hours per paper, in comparison to the average of 23 hours per paper from GTC open-time proposals (Cabrera-Lavers, 2020). We note that MIR observations are inherently less efficient than those in the visible/near-IR wavelengths due to the overheads of the chop/nod technique. We also note MIR demand from the GTC community represents a small fraction of the telescope overall production, more biased toward optical/near-IR wavelengths.

## 1.2 On the Chop/Nod Technique

As noted in §1.1, MIR observations typically require the secondary mirror of the telescope to be oscillated at a frequency of a few times a second to subtract the background and sample the noise. The frequency of the chopping (for the current generation of MIR arrays) is typically set by the array so-called (very fast changing) "1/f" noise, but also removes the time-variable sky background (fast changing, i.e. Pantin, 2010) and telescope thermal emission (slow changing). In the *N* band (~7.7-13μm) at a good site and under good conditions an approximate estimate for the characteristic sky emission stability is ~0.2 to 20 seconds (dependent on the wavelength and exact conditions, Pantin, 2010) and the telescope thermal emission stability is >~20 seconds. Array stability is highly dependent on the array, array temperature, and applied voltages array, but in the case of the CanariCam Raytheon CRC774 array, a chop frequency of a few Hertz effectively removed the 1/f noise.

Chopping the secondary mirror induces several deleterious effects to the data and data collection efficiency including:

1. Fast guiding is often available in only one beam of the chopped image, sometimes rendering the unguided beam unusable due to the degraded point spread function (PSF). Thus, the photon collection time is reduced by this single effect by 50%, or an often greatly inferior PSF must be accepted.

2. Reduced on-source photon collection time, as the mirror is physically moving and settling to a stable position for a fraction of the elapsed (clock) time.

3. As the beam propagates through the telescope through two slightly different optical paths, the telescope emission is not precisely the same in the frame of reference of the array, leading to a so-called radiative offset (i.e. Burtscher et al., 2020). This can be countered by nodding the telescope in an equal and opposite amount of the chop and repeating the observations in this 'negative' beam. The nodding cadence is typically a few 10's of seconds, dependent on the rotation rate of the pupil and the centrosymmetry of the pupil, which can also imprint a radiative offset on the final data product. As the pupil rotation rate is a function of zenith angle, a canonical value is often established during commissioning which is a compromise between minimization of the number of nods and accurate radiative offset removal. The time required for nodding and subsequent settling further reduces the photon collection time.

4. As the secondary mirror is partially de-collimated by the act of chopping, optical aberrations are induced in the chopped beam, increasing in severity as a function of chop distance. This is shown in Figure 1 for the GTC/CC system.

5. It is possible that the off-source (chopped) beam can contain other sources and potentially extended sources emission, therefore resulting in incorrect sky subtraction.

---

[1] As measured during the commissioning of CanariCam and described here: http://www.gtc.iac.es/instruments/canaricam/canaricam.php#ImageQuality

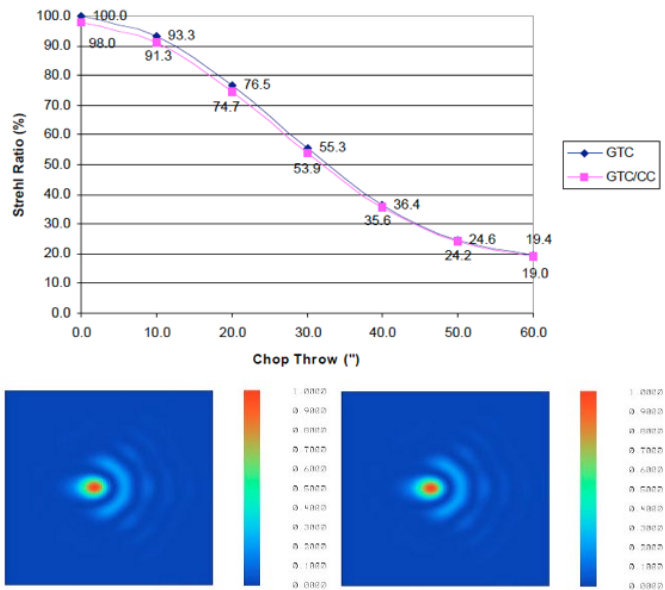

**Figure 1.** Model effect of the chop throw on CanariCam Strehl ratio (top). Models of the image quality delivered by the GTC (lower left) and CanariCam/GTC (lower right) for a chop throw of 30", shown in log-scale.

These effects combine to place tight constraints on MIR ground-based observations, and the reduction in on-source photon collection time is typically significant. In the case of the GTC/CanariCam, and where the chopped beam is excluded (as the lack of guiding resulted in a significant PSF degradation), the observing efficiency (defined as photon collection time/clock time) is ~32%, typical for MIR ground-based observations at, for example, the GTC, Gemini, UKIRT, and other 4-10m observatories. If the unguided chopped beam (the 'negative' beam) is included to the guided beam (the 'positive' beam), the observing efficiency doubles to ~64%, but this is not commonly the case as the unguided beam typically has an inferior PSF which serves to (a) reduce the final image quality and thus (b) require a larger photometric aperture which includes more pixels which themselves have a noise profile which can diminish or reverse the increased signal to noise ratio (SNR) due to the increased noise of the larger photometric extraction aperture. When observing in the MIR on an 8 m-class telescope, obtaining optimal image quality is typically a key science driver and so in our experience it is rare to include the unguided negative beam.

Finally, the issues listed above do not pay any regard to the demanding engineering requirements that a rapidly oscillating mirror imposes on the telescope. Indeed, for the 30m class telescopes, the secondary mirror will be too large to chop (i.e. in the case of the TMT, M2 is 3.1m is diameter) and hence other methodologies for removal of the array/sky noise are required. Instead, promising work proceeds for chopping/nodding a mirror located at a pupil plane interior to the instrument (i.e. Honda et al. 2020), principal component analysis (PCA, Hunziker et al., 2018), and so-called drift scanning techniques (i.e. Ohsawa et al., 2018, PTF 2021).

*1.3 Drift Scanning and MIR Arrays*

Drift scanning makes use of the MIR array capabilities to be rapidly readout whilst the telescope is either held at a single position with respect to the mount (i.e. not tracking the projection of the moving sky) or being driven at a user defined speed across the sky field (i.e. tracking plus/minus an increment in R.A. or Dec). Thus, each image contains a short exposure (measured in units of ms) which can be differenced from an earlier or subsequent obtained frame to correct for the sky background and then shifted and co-added in subsequent data reduction. The secondary remains fixed, so the observing efficiency rises to ~100%, and optical aberrations are eliminated as compared to chopping, as are guiding issues in the chopped beam. However, unless the drift scan speed is fast (greater than a few seeing discs per second), noise subtraction will be inferior or require more advanced data reduction techniques, such as PCA or other noise mitigation strategies. Conversely, if the drift scanning is too fast the image of the object will be elongated along the scanning direction hence an optimal balance must be established.

CanariCam uses a CRC774 Raytheon array as the MIR sensor. Consisting of 320x240 pixels, it was state of the art at the time of CanariCam's design. However, it is well known that this array suffers from several array artifacts due to the IR sensitive materials, multiplexor used, and readout approach (Sako et al. 2003). CanariCam utilizes a readout technique to mitigate those effects, but residuals effects remain. In standard chop/nod mode, the primary residual effect is a "cross-talk" in each of the 16 channels of 20x240 pixels width, where a bright object in one channel causes a negative image in each of the other 15 channels, but at a fraction of the object flux. The inverse is true for the chopped beam, where the negative beam creates a positive cross-talk effect. When drift scanning, this effect is also present but at a lower level as specific pixels receive less photons than the chop/nod pixels typically receive. In this paper, we compare the results of drift scanning of a photometric standard with those made in standard chop/nod mode, temporally separated by at most a few hours.

**2. Observational Approach and Data Collection**

*2.1 Telescope Setup*

While drift-scanning is not an observing mode offered by GTC, standard non-sidereal tracking and guiding functionality is flexible enough to carry out a drift-scanning observation with minimal effort. By creating an artificial ephemeris file for our sidereal object, specifying virtual positions in astrometric right ascension and declination coordinates, and velocities at tabulated time intervals, we can drift the object at the desired rate following an arbitrary path across the FOV, as shown in Figure 2. The main advantage of this highly flexible approach is that these technical observations did not require any modification to the GTC Control System (GCS, Filgueira & Rodriguez, 1998) but still afford the same results as true drift scanning.

The operational drawbacks were related to managing the synchronization between the telescope non-sidereal tracking and the acquisition process on the instrument field of view, since, ideally, one would like to begin the drift from a specific known position on the detector when the detector starts exposing. Also, the ephemeris file must be generated ad-hoc for the target and for a given time slot because the drift is to be performed on sky R.A. and DEC coordinates, and we created a Python script to automate this process. For future development, a drift pattern could be programmed at the GCS-level using the focal plane coordinate system (fixed with respect to the instrument), which would make it more generally applicable to any target on-sky and easier to synchronize with the instrument acquisition.

## 2.2 Observation Description and Log

To compare the SNR obtained in imaging mode, we made observations of the same standard star in (1) chop/nod mode (a north-south pattern separated by 5.7", which is smaller than the typical 7-15", to provide the optimal PSF whilst ensuring the chopped beams are well separated) and (2) drift scanning mode, temporally separated by only a few minutes. Assuming background stability across the observing time, this enables comparison of those data sets and hence characterizes the difference in SNR for both (a) on-source exposure time and (b) clock elapsed time. For this experiment, we define the background to be the addition of sky, telescope, window, array, and all other sources.

The source chosen was of sufficient brightness to obtain excellent SNR in both a short chop/nod exposure, as well as individual images within the drift scan stack. When obtaining the drift scan data, the telescope was driven in non-sideral tracking at 3 different speeds. In all cases, the star was moved to describe a rectangular shape across the 320x240 array, because it allows us:

1. To characterize performance for the two drift directions independently (whether drifting along the detector channels or across them have any impact in the final instrument sensitivity due to the detector artifacts), and

2. To minimize the synchronization requirements between the telescope drift and the instrument integration because the target will always be in the FOV and following the same path for a long period of time.

We drove the drift scan at 3 different speeds to empirically evaluate which is the optimal speed from SNR and FWHM metrics for our given environmental conditions and instrumental setup. However, in this paper we describe only the fastest drift scan speed, a comparison of the three speeds was made in PTF 2021. The speed of drifting was set to be a multiple of the seeing disk, and through this experiment we try to balance the degradation of the FWHM vs. the improvement of obtaining better sampled background subtraction.

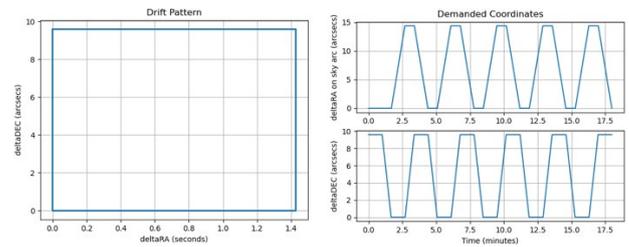

**Figure 2.** Representation of the ephemeris file and demanded astrometric coordinates for one of our observations.

Due to limitations in the data acquisition system, the detector controller cannot transfer all individual frames, with typical integration times of 25 ms, but it must accumulate a certain number of them in its internal buffer, and finally transfer and save an accumulated image, called "saveset", at a maximum rate of approximately 5 Hz (savetime=0.207s). This constrains the maximum drift rate at which the PSF will not be elongated in each saveset, and we chose the maximum drift as 3 pixels sec$^{-1}$, which degraded the PSF a fraction of a pixel. This is the main limiting factor of CanariCam drift-scan mode because this ultimately limited the frequency at which we could subtract the variable background.

MIR median conditions at the Observatorio del Roque de Los Muchachos imply a FWHM ~0.32", which corresponds to 4 pixels on the CanariCam array, with a plate scale of 0.08" pixel$^{-1}$. With a 3 pixel sec$^{-1}$ drift, the PSF moved one seeing disk every 1.33 seconds, while with a standard chopping configuration of 2Hz, the background was subtracted every 0.25 seconds, better sampling its variations along the observation. The full observing log and comments regarding the precipitable water vapor are made in Table 1.

## 3. Data Reduction

We use and advance the approaches and results detailed in PTF 2021, and in this section we discuss the similarities and contrast the differences, and later improvements in our updated data reduction process[2].

### 3.1 Drift Scan Data Reduction

As described by PTF 2021, we divide the background counts ($B$) into five sources, the sky ($S$), telescope thermal emission ($T$), the instrument background ($I$, including entrance window), array bias ($A$), and array/readout noise ($N$). As previously noted (section 1) each of these parameters vary at their own characteristic frequency. The object ($O$) was drift scanned at three different rates across the array (section 2). A single frame at a given frame number ($F_n$, where $n$ is the frame at number ($n$) = 1-2,901) consists of the object ($O_n$) plus the background unique to that frame ($B_n = S_n+T_n+I_n+A_n+N_n$). To optimally reduce the drift scan data we computed $B_n$ at the fastest cadence appropriate to minimize variations in $B_n$. Whilst we include $N_n$ within $B_n$, as we later show, it is necessary to account for this component in an independent and secondary reduction step.

---

[2] The data reduction pipelines are accessible through the following link: https://github.com/Locuan/canaricam-pipelines

To produce an estimate for $B_n$, our approach was to produce a rolling median sky of five images separated by an increment of $\delta n$. As the object is moving across the array, a median frame effectively removes the object from the resultant median frame. In PTF (2021) we found for the fastest drift scan $\delta n=26$ ($\delta t=5.4s$) was sufficiently spatially separated that differenced imaged affected the photometry by <5%, and hence we adopted this metric for our median sky $\delta n$. We found for a median of three frames the source was not well removed, possibly due to frame persistence, but five frames were sufficient. The median frame is thus a rolling median that is the minimum temporal separation whilst ensuring negligible overlapping of the object images. The median background frame ($M_n$) can be described as:

$$M_n = \mathrm{median}[F_{n-52}, F_{n-26}, F_n, F_{n+26}, F_{n+52}]$$

In the extreme cases at the start and end of the drift scan (i.e. $n=1$), $M_n$ is forced to be estimated at more temporally distant frames (i.e. $F_1, F_{27}, F_{53}, F_{79}, F_{105}$) until $n=53$ where upon the approach above can be followed.

For each object frame $O_n$, we then subtracted the $M_n$ frame to produce a background ($B_n$) corrected image. However, as noted before, the array/readout ($N_n$) is not accounted for in this step as the variation of $N$ is on a much faster cadence than the construction of the $M_n$ frame. Thus, we now have $O_n+N_n$. We next produce a noise estimate for *each* frame to avoid the $M_n$ cadence problem.

To produce a noise frame, $N_n$, for each frame we first removed the source from the frame by setting a 28 x 28 pixel box centered at the centroid of the object to zero counts. This box size was set to be ten times the final measured FWHM of the combined data, thereby ensuring the source was completely removed but minimizing the number of pixels that were set to zero. Next, the image was Fourier transformed to produce an image of the spatial frequency contained in $O_n+N_n$. The object was removed to ensure no frequencies associated with the object were produced in this Fourier transform. In the case of CanariCam, a clearly distinct frequency occurring at every 16 pixels was detected in Fourier space, totaling 20 "tramlines." We suggest this spatial frequency is due to the format of the array and its readout electronics, with 16 channels of 20 pixels each, replicating a similar (bias and gain variation) structure in each channel. Next, we isolated these tramlines by setting all other pixels to zero and transformed this resultant of the tramlines only image back to image space. This represents $N_n$, which was then subtracted from the residual of $F_n$ after $M_n$ subtraction to give the final frame $O_n$, as described below:

$$O_n = F_n - M_n - N_n$$

The object is next extracted from the $O_n$ frame in a 80 x 80 pixel box (as the object was ≥40 pixels from the array edge during the drift scan) centered on the object. The process was repeated for each frame within the drift scan, and finally the 80 x 80 pixel frames were coadded to form a final image. The entire data reduction process is shown graphically in Figure 3.

*3.2 Chop/Nod Data Reduction*

To reduce the chop/nod data, we followed four methodologies. The first was a standard chop/nod data reduction, as described in multiple other MIR papers. The second was to extract similar sized boxes centered on the positive and two negative beams, invert the negative beams, and co-add the three images. In this second methodology the effective on-source time was doubled at the expense of including the unguided and hence inferior PSF images. The third methodology (Figure 4) was a modified reduction approach leveraged from the first standard chop/nod method. In this methodology, within each nod position we firstly co-added all the on- and off-source beams, and then differenced those results. We then removed the positive and negative source images from the frame by setting a 34 x 34 pixel box centered at the centroid of the object to zero counts. This box size was set to be ten times the final measured FWHM of the combined data, thereby ensuring the source was completely removed but minimizing the number of pixels set to zero. As with the drift scan data, the image was Fourier transformed, where a clearly distinct frequency occurring at every 16 pixels was detected. As before, we isolated these tramlines by setting all other pixels to zero and transformed this resultant image back to image space. This represents $N_n$ for the chopped data, which was then subtracted the residual image after chop subtraction and is the smallest temporal quanta that the noise frame can be achieved. This was then repeated for each nod position and the radiative offset removed by differencing those nod positions. The differenced nod positions were combined to produce a final image. The object is next extracted in a 70x70 pixel box (the positive and negative beams were separated by ~72 pixels thus we extracted to 50% of the distance between those beams) centered on the object. Finally, for the fourth reduction, we used the noise corrected frames from method three, and then followed the approach of method two where the guided and unguided beams were co-added.

Thus, the four types of data reduction results in four different finally reduced final images: (1) standard chop/nod, (2) co-added beams, (3) chop/nod with noise subtraction, (4) chop/nod with noise subtraction and co-added beams.

*3.3 SNR Estimations*

To characterize, compare, and contrast the results of the data colletion/reduction methodologies, we measured the (a) the noise in the sky, far from the object, (b) signal of the source using aperture photometry, (c) signal of the source using optimal photometry, (d) the FWHM of the image, and (e) the Strehl ratio.

The noise (a) was estimated through an average rms in four regions of 10x10 pixels in the corners of the extracted data frame, where the center of the noise boxes were 49.50 pixels from the source centroid for the drift scan. In the case of the chop/nod the center of the noise boxes were 42.43 pixels from the source centroid. The signal (b) in a fixed aperture size was then measured. We followed the approach of PTF (2021) and used an aperture radius which encircled >95% of the flux of the drift scan source, and determined this as our fixed aperture size. We then measured the signal for each of our results, using the same sized aperture. Similarly, we determined the aperture size required to enclose >95% of the flux for our chop/nod data and used that fixed aperture size to measure the signal for each of our results. For the optimal photometry (c) we used the results from our fixed aperture determination to compare our photometric results where the aperture enclosed 95% of the flux for each of the data collection/reduction techniques. The FWHM (d) was then estimated using standard techniques, and finally the Strehl ratio was estimated by comparing the peak of the normalized PSF measured in the data divided by the peak of a normalized PSF in the theoretical PSF images computed from the pupil of the GTC sampled at the same plate scale as CanariCam and the wavelength of observations. In the next section we tabulate and discuss the results.

## 4. Results

Figure 5 shows the final images resultant from the data reduction approach as described above, and Figure 6 shows the chop/nod data both with and without noise subtraction to demonstrate the improved background. Table 2 shows the FWHM of the source for the drift scan, the three chop/nod methods discussed above, and also the results shown in PTF (2021). We use the CanariCam plate scale of 0.0798 arcseconds per pixel to convert to angular size, and finally include the diffraction limit using $\lambda/D$ for comparison at the filter central wavelength (8.7μm) and the telescope diameter of 9.4m due to the use of an inscribed pupil stop interior to CanariCam. In the same table, we quote the Strehl ratio. We note the improvement in the drift scan data as compared to the more traditional chop/nod approach, attributable to an approximation of post-processing shift/add, and speculate that a faster frame rate could improve these values futher. Indeed, the Strehl ratio for the drift scan is (in our experience) superior to other data we have used, where 3 Airy rings are clearly detected, and sections of a fourth ring is tentatively evident. Finally, the FWHM is modeled as a simple Gaussian for all datasets. As shown by Radomski et al. (2008), near the diffraction limit in the MIR a Moffat profile more closely resembles the image profile and hence the reported Strehl ratio is likely a lower limit and likely accounts for the slight difference in the drift scans reduction (this paper compared to PTF 2021). The chop/nod readout approach to not self-cancel the source (i.e. to not add the positive and negative source) for CanariCam, and limitation of its readout electronics, removes the ability to perform high temporal cadence shift/add. The effect of shif/add on the avaialble data is marginal (<<1%), as shown in PTF 2021.

In Tables 3, 4, and 5 we show the SNR estimates for the aperture sizes set by the drift scan FWHM, chop/nod FWHM, and optimal respectively. Table 3 shows includes the results from PTF 2021 drift scan reduction which did not include noise estimation/subtraction, as well as the chop/nod reductions. We note the very substantial increase in SNR for this paper's drift scan reduction, showing the importance of both the noise estimation/subtraction as well as the improvements associated with the benefits of drift scanning. We also note that signal counts are reduced by 1.05% when comparing the counts in this paper to the PTF 2021, but in the latter 26 of the 2,900 frames are not used due to the specifics of the data reduction, corresponding to 0.90% of the total. Thus the signal count difference is insignificantly different. Further, as the count rate in the two chop/nod reductions are similar, the noise subtraction approach has a negligible effect to the signal count rate and thus photometric integrity is preserved using this methodology.

Table 4 shows a similarly substantial increase in the SNR, but the most significant improvements are shown in Table 5 where the improved FWHM of the drift scan data is fully exploited where a smaller aperture size enclosed >95% of the flux compared to the chop/nod data. This leads to a reduced residual noise profile, and the large increase in the derived SNR where comparing standard chop/nod data to the drift scan data (more than a factor of 3). Table 6 shows the clock time (i.e. telescope time) and on-source photon collection time for the drift scan and chop/nod. We contend that the clock time is the correct metric to compare the datasets, as this is the increment that time allocation committees award to successful programs. The difference in clock time is ~3%, with a slightly longer time provided to the chop/nod approach. If comparing the on-source times, the drift scan is a factor of 3.02 times longer exposure time, which using a simple scaling of SNR $\alpha$ $t^{0.5}$ (where t is the on-source time), the SNR should be ~74% higher. The SNR comparison between the drift scan and chop/nod standard data reduction in all cases shows a substantially higher increase than 74% and thus extra photon collection time is not the unique contributing factor to the improved SNR. This conclusion is further supported when comparing the chop/nod reductions, where the noise subtraction is shown to be much more significant in the SNR improvement compared to co-adding the guided and unguided beams.

In summary, comparing the drift scan and chop/nod results, for the quantities of FWHM, Strehl ratio, and SNR, in all cases the driftscan approach affords signficant improvements.

## 5. Conclusions and Discussion

In this paper we have presented the results of drift scanning and the substantial improvements it affords to specific CanariCam observations, namely a point source in an uncrowded field. This is a special case, and may not be generally realized for other observations, such as those of extended objects, crowded fields where the drifting may not be possible, or indeed to other instruments. When comparing the SNR where noise subtraction was not used to data where it was, there is a significant improvement. However, a substantially higher SNR improvement was found for the drift scan compared to the chop/nod data, presumably due a combination of (1) the extra photon collection time and (2) the improved estimation of the background (compared to chop/nod observations) which we suggest would be likely commutable to other MIR instruments. We speculate that the improved background estimation at least partially arises through the median of 5 spatially and temporally distinct frame compared to a single frame used in chop/nod. Further, the improvement to the final PSF permits a smaller photometric extraction aperture to be used which is also commutable to other telescopes/instruments.

The sensitivity of James Webb Space Telescope (JWST) will usher in a new era of IR astronomy, and whilst the PSF will be far more stable than ground-based telescopes, the FWHM will be inferior to the 8-10m class of telescopes. Instruments such as the proposed MICHI MIR imager/spectromer on the the TMT (Packham et al., 2012) will boast a PSF as small as 55 mas, and this equisite spatial resolution (i.e. Nikutta et al. 2021a, 2021b) and planned very high spectral resolution will be highly complimentary to the JWST sensitivity. At least some of the key MICHI science cases are of objects that are extended along a plane (i.e. protoplanetary disks) that could be oriented to enable drift scanning observations and the improved SNR and PSF, or where there is a companion source (i.e. an exoplanet) which could also be oriented to permit the benefits of drift scanning to be realized. We also note that for observations of low and geosynchronous Earth orbit (LEO, GEO) objects which have lage non-sideral motions, a form of drift scanning where the telescope siderally tracks but the LEO and GEO objects drift through the field of view could be employed.

The CanariCam array and readout electronics are now redundant, and likely the next generationa of MIR instruments will use the Teledyne Geosnap array (i.e. Bowens et al., 2020). The precise noise and operational characterstics of this array are being characterized, but we note the rapid readout of the array is potentially especially useful in a drift scan observations. The larger format of the Geosnap (up to 3,000 x 3,000 pixels) allows for far greater flexibility in drift scan patterns and exposure lenghts. CanariCam is now decommissioned at the GTC, preventing experiments that we would very much like to conduct: (1) changing the drift scan speed to optimize the final SNR and (2) characterize spectrographic drift scanning.

In summary, our data shows the promise of drift scanning, which additionally eliminates the need for chop/nod mechanisms and associated overheads.

## 6. Acknowledgements

Based on observations made with the Gran Telescopio Canarias (GTC), installed at the Spanish Observatorio del Roque de los Muchachos of the Instituto de Astrofísica de Canarias, in the island of La Palma. The upgrade of CanariCam was co-financed by the European Regional Development Fund (ERDF), within the framework of the "Programa Operativo de Crecimiento Inteligente 2014-2020", project "Mejora de la ICTS Gran Telescopio CANARIAS (2016-2020)". The authors wish to acknowledge the help of the entire GTC staff (science and engineering) as well as those key people in the CanariCam project, originated at the University of Florida. ARTQ wishes to acknowledge John Kucewicz and Sharon Smith-Kucewicz for their generous financial assistance received in support of his graduate studies. ARTQ also acknowledges support from the Department of Physics & Astronomy at UTSA. We wish to thank the anonymous referee for comments that helped to improve this manuscript.

**Table 1**
Log of observations

| Date | UTC at start | Elapsed time (s) | On-Source Time (minutes) | Savetime (s) | Object | Filter | Observing Mode / Setup | Comments |
|---|---|---|---|---|---|---|---|---|
| 28/07/2020 | 01:04:00 | 619.55 | 3.31 | 5.79 | HD213310 | Si2-8.7 | Chop-Nod 5.7 arcsec North (up) Chop Freq. = 2.07Hz | Reference data before drift scan experiment to measure sensitivity. Good image quality. PWV=6.5 mm |
| 28/07/2020 | 01:31:03 | 600.06 | 10.001 | 0.207 | HD213310 | Si2-8.7 | Stare. Drift rate = 0.24 arcsec/s | Drift scanning #1 at fast rate of ~3 pix/s. PWV between 6.3 and 5.7 mm |
| 28/07/2020 | 01:51:13 | 600.06 | 10.001 | 0.207 | HD213310 | Si2-8.7 | Stare Drift rate = 0.08 arcsec/s | Drift scanning #2 at medium rate of ~1 pix/s. PWV=6.2 mm |
| 28/07/2020 | 02:02:18 | 600.06 | 10.001 | 0.207 | HD213310 | Si2-8.7 | Stare Drift rate = 0.08 arcsec/s | Repeat. PWV=6.2 mm |
| 28/07/2020 | 02:17:03 | 600.06 | 10.001 | 0.207 | HD213310 | Si2-8.7 | Stare Drift rate = 0.04 arcsec/s | Drift scanning #3 at slow rate of ~0.5 pix/s. PWV=5.8 mm |
| 28/07/2020 | 02:27:14 | 600.06 | 10.001 | 0.207 | HD213310 | Si2-8.7 | Stare Drift rate = 0.04 arcsec/s | Repeat. PWV between 5.8 and 6.0 mm |
| 28/07/2020 | 02:42:32 | 401.49 | 2.137 | 1.45 | HD213310 | Si2-8.7 | Chop-Nod 5.7 arcsec North (up) Chop Freq. = 2.07Hz | Reference data after drift scanning. Good image quality. PWV between 6.0 and 6.3 mm |

**Figure 3**
Drift Scan Data Reduction Flow Chart

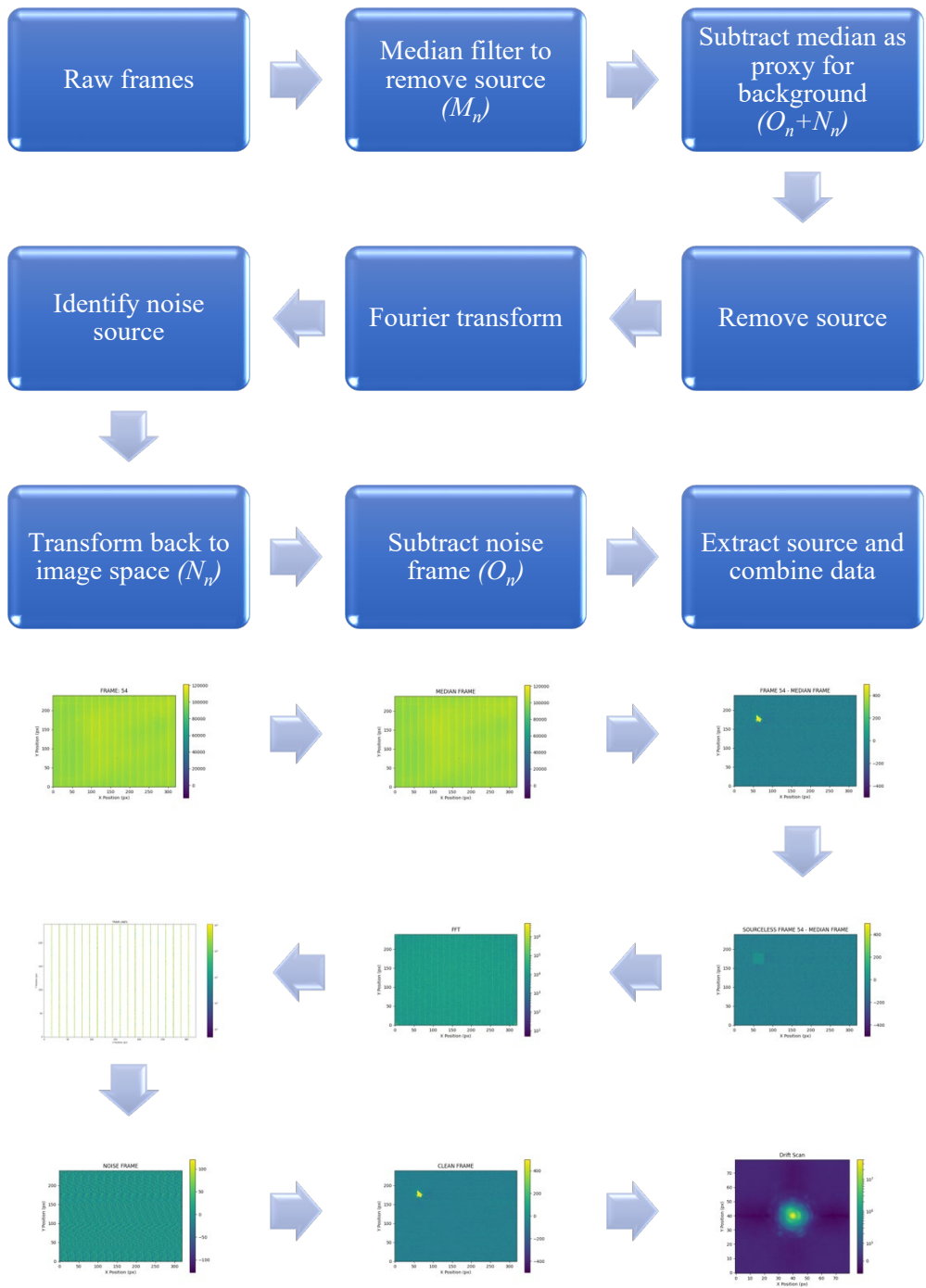

**Figure 4**
Modified Chop/Nod Data Reduction Flow Chart

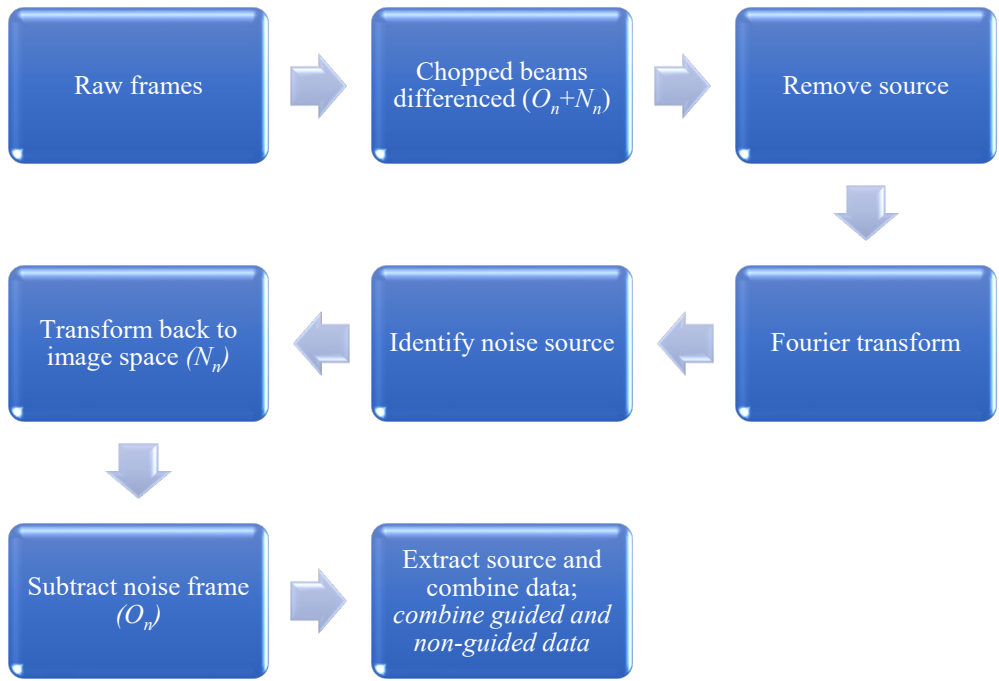

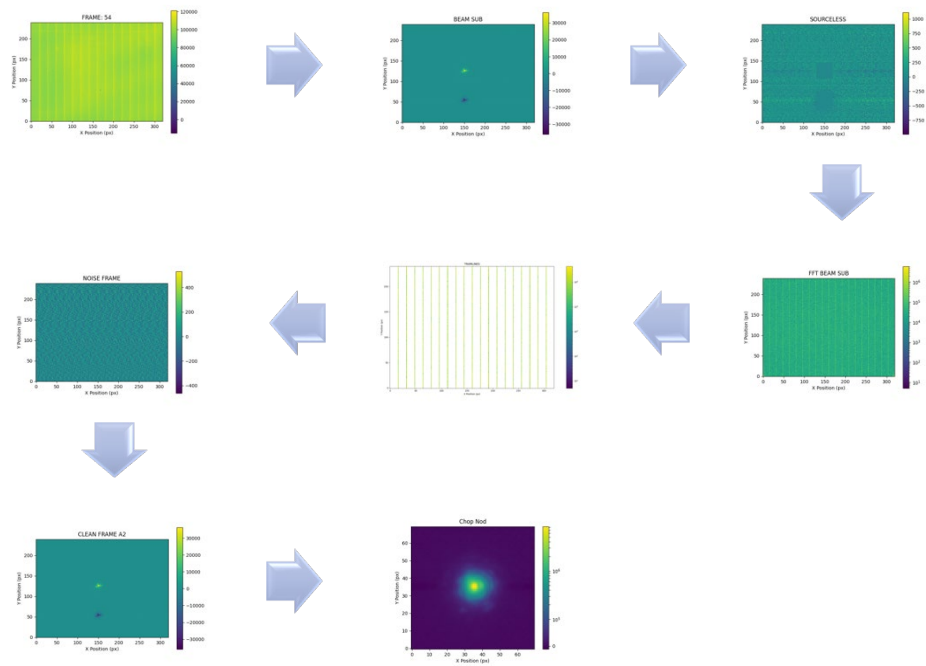

**Figure 5**
Final drift scan (left) and chop/nod (right) after noise estimation/subtraction images

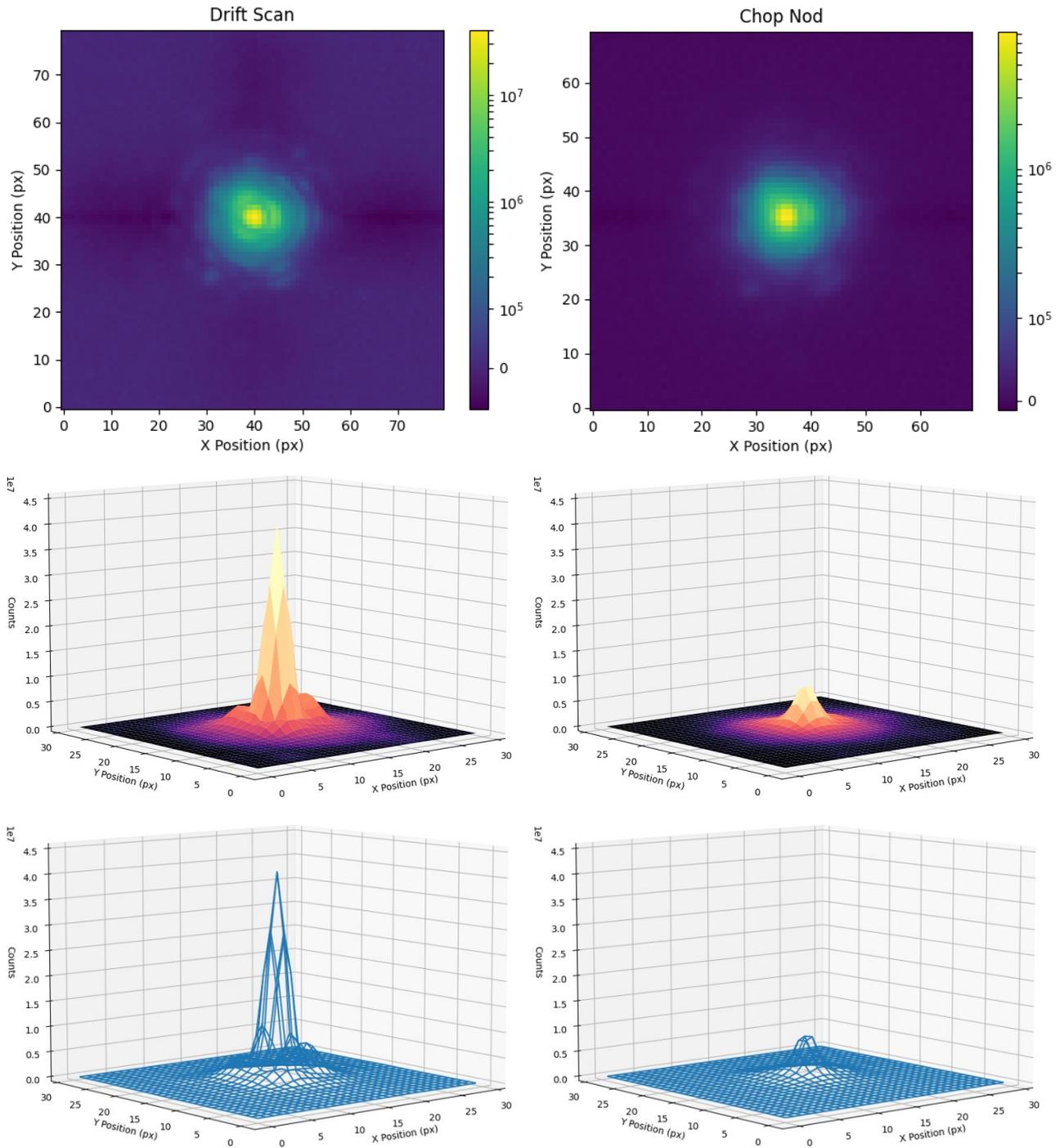

**Figure 6**
Detail of the noise residuals in the chop/nod data without noise estimation/subtraction (left) and after (right).

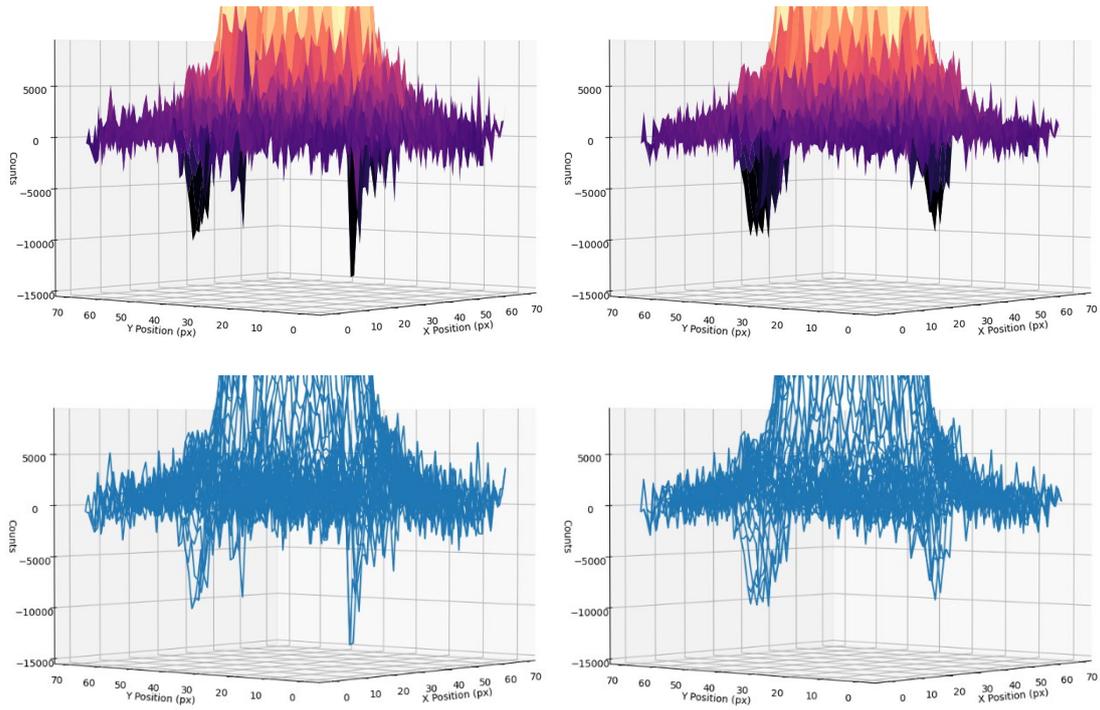

**Table 2**
FWHM and Strehl ratio for the various data collection/reduction methodologies

| Data Collection/Reduction | FWHM: pixels, " | Strehl Ratio |
|---|---|---|
| Drift scan (this paper) | 2.80, 0.223 | 54% |
| Drift scan (PTF 2021) | 2.80, 0.223 | 57% |
| Chop/nod (standard reduction) | 3.32, 0.264 | 35% |
| Chop/nod (noise subtraction) | 3.32, 0.264 | 35% |
| Chop/nod (co-added beams) | 3.37, 0.269 | 34% |
| Chop/nod (noise subtraction & co-added beams) | 3.37, 0.269 | 34% |
| *Diffraction limit* | *2.39, 0.191* | *100%* |

**Table 3**
SNR estimates for the aperture size set by the drift scan FWHM: Δ SNR is the change in SNR from the drift-scan data newly reduced (this paper).

| Data Collection/Reduction | Aperture Radius (pixels) | Noise per pixel (counts) | Aperture Signal (counts) | SNR | Δ SNR |
|---|---|---|---|---|---|
| Drift-scan (this paper) | 8.66 | $2.209 \times 10^3$ | $6.136 \times 10^8$ | $1.789 \times 10^4$ | - |
| Drift scan (PTF 2021) | 10.00 | $3.700 \times 10^3$ | $6.072 \times 10^8$ | $9.396 \times 10^3$ | 90.4% |
| Chop/nod (standard reduction) | 8.66 | $1.650 \times 10^3$ | $2.036 \times 10^8$ | $7.947 \times 10^3$ | 125.2% |
| Chop/nod (co-added beams) | 8.66 | $2.805 \times 10^3$ | $4.054 \times 10^8$ | $9.310 \times 10^3$ | 92.2% |
| Chop/nod (noise subtraction) | 8.66 | $1.251 \times 10^3$ | $2.036 \times 10^8$ | $1.049 \times 10^4$ | 70.7% |
| Chop/nod (noise subtraction & co-added beams) | 8.66 | $2.180 \times 10^3$ | $4.056 \times 10^8$ | $1.199 \times 10^4$ | 49.3% |

**Table 4**
SNR estimates for the aperture size set by the chop/nod FWHM: Δ SNR is the change in SNR from the drift-scan data newly reduced (this paper).

| Data Collection/Reduction | Aperture Radius (pixels) | Noise per pixel (counts) | Aperture Signal (counts) | SNR | Δ SNR |
|---|---|---|---|---|---|
| Drift-scan (this paper) | 15.39 | $2.209 \times 10^3$ | $6.433 \times 10^8$ | $1.064 \times 10^4$ | - |
| Chop/nod (standard reduction) | 15.39 | $1.650 \times 10^3$ | $2.278 \times 10^8$ | $5.045 \times 10^3$ | 110.9% |
| Chop/nod (co-added beams) | 15.39 | $2.805 \times 10^3$ | $4.534 \times 10^8$ | $5.906 \times 10^3$ | 80.2% |
| Chop/nod (noise subtraction) | 15.39 | $1.251 \times 10^3$ | $2.276 \times 10^8$ | $6.649 \times 10^3$ | 60.0% |
| Chop/nod (noise subtraction & co-added beams) | 15.39 | $2.180 \times 10^3$ | $4.532 \times 10^8$ | $7.597 \times 10^3$ | 40.1% |

**Table 5**
SNR estimates for the optimal aperture size: Δ SNR is the change in SNR from the drift-scan data newly reduced (this paper).

| Data Collection/Reduction | Aperture Radius (pixels) | Noise per pixel (counts) | Aperture Signal (counts) | SNR | Δ SNR |
|---|---|---|---|---|---|
| Drift-scan (new) | 8.66 | $2.209 \times 10^3$ | $6.136 \times 10^8$ | $1.789 \times 10^4$ | - |
| Drift scan (PTF 2021) | 10.00 | $3.700 \times 10^3$ | $6.072 \times 10^8$ | $9.396 \times 10^3$ | 90.4% |
| Chop/nod (standard reduction) | 15.39 | $1.650 \times 10^3$ | $2.278 \times 10^8$ | $5.045 \times 10^3$ | 254.7% |
| Chop/nod (co-added beams) | 15.98 | $2.805 \times 10^3$ | $4.534 \times 10^8$ | $5.906 \times 10^3$ | 202.9% |
| Chop/nod (noise subtraction) | 15.39 | $1.251 \times 10^3$ | $2.276 \times 10^8$ | $6.649 \times 10^3$ | 169.2% |
| Chop/nod (noise subtraction & co-added beams) | 15.98 | $2.180 \times 10^3$ | $4.549 \times 10^8$ | $7.410 \times 10^3$ | 141.4% |

**Table 6**
Photon Collection Time

| Method | Clock Time (s) | On-Source Time (s) | Observing Efficency (%) |
|---|---|---|---|
| Drift scan | 600.06 | 600.06 | 100 |
| Chop/nod (standard) | 619.55 | 198.6 | 32 |
| Chop/nod (co-added beams) | 619.55 | 397.2 | 64 |